# Observation of degenerate one-dimensional sub-bands in cylindrical InAs nanowires

Alexandra C. Ford[1,2,3,†], S. Bala Kumar[4,†], Rehan Kapadia[1,2,3], Jing Guo[4], Ali Javey[1,2,3,]*

[1] *Electrical Engineering and Computer Sciences, University of California, Berkeley, CA, 94720, USA.*

[2]*Materials Sciences Division, Lawrence Berkeley National Laboratory, Berkeley, CA 94720, USA.*

[3]*Berkeley Sensor and Actuator Center, University of California, Berkeley, CA, 94720, USA.*

[4] *Electrical and Computer Engineering, University of Florida, Gainesville, Florida 32611*

[†]These authors contributed equally.

* Corresponding author: ajavey@eecs.berkeley.edu

**ABSTRACT** – One-dimensional (1D) sub-bands in cylindrical InAs nanowires (NWs) are electrically mapped as a function of NW diameter in the range of 15-35 nm. At low temperatures, stepwise current increases with the gate voltage are clearly observed and attributed to the electron transport through individual 1D sub-bands. The two-fold degeneracy in certain sub-band energies predicted by simulation due to structural symmetry is experimentally observed for the first time. The experimentally obtained sub-band energies match the simulated results, shedding light on both the energies of the sub-bands as well as the number of sub-bands populated per given gate voltage and diameter. This work serves to provide better insight into the electrical transport behavior of 1D semiconductors.

One-dimensional (1D) materials such as semiconductor nanowires (NWs) have been employed in a variety of nanoelectronic applications such as field-effect transistors (FETs), light-emitting diodes (LEDs), lasers, and sensors.[1,2,3,4,5,6,7] Consequently, a more thorough understanding of the fundamental physics of 1D systems is essential. Indium arsenide (InAs) is a particularly interesting material for the study of quantum phenomena as it has a large Bohr radius of ~34nm, resulting in strong quantum confinement effects clearly observable in relatively large diameter NWs.[8,9,10,11,12,13] Additionally, due to Fermi level pinning in the conduction band at the metal interfaces, ohmic contacts can readily be formed to InAs.[14,15] Furthermore, the surface of InAs can be passivated with $ZrO_2$ to enable low density of surface trap states, thereby, enabling detailed characterization of the intrinsic carrier transport properties of InAs nanostructures.[13,16] While simulations of 1D sub-band energy calculations appear in the literature, there are few studies that carefully match simulation to experimental results, especially as a function of NW diameter.[17,18,19,20,21,22,23] In this paper, we electrically probe the sub-band energies of InAs NWs for a diameter range of 15-35 nm through experiment and simulation. Interestingly, the two-fold degeneracy in certain 1D sub-band energies arising from structural symmetry is experimentally observed in cylindrical NWs.

A schematic illustrating the cross-sectional structure of a representative back-gated InAs NW device is shown in Figure 1a. The InAs NWs used in this study were synthesized using the vapor-liquid-solid mechanism with Ni nanoparticle catalysts by vapor transport technique using solid powder InAs source as previously described.[24] A ~2.5nm native oxide shell is present around the NWs, as previously observed by high resolution transmission electron microscopy.[15,24] NWs grown by this method typically have InAs diameters $d$=14-36nm post-native oxide subtraction. To fabricate devices, the InAs NW growth substrates were placed in anhydrous ethanol, the InAs NWs detached from the growth substrates by sonication, and dropcast on 50nm thermally grown $SiO_2$/Si substrates. The $p^+$ Si substrate is used as the global back-gate. The source (S) and drain (D) contacts were defined via photolithography and thermal/e-beam evaporation of Ni/Au, with a ~5s treatment in 50:1 HF to remove the native oxide immediately prior to metallization. The fabricated field-effect transistors (FETs) had channel lengths $L$=6-12μm. The InAs NW surfaces were then passivated with 7nm $ZrO_2$ deposited by atomic layer deposition at 130°C and 1.1μm OiR 10i (I-line) photoresist (PR). This surface passivation is important for observing the clear electron population of 1D sub-bands as a function of gate voltage. The $ZrO_2$ surface passivation reduces the interface traps[16] and surface disorder. While the interface traps are frozen out at 77K as previously reported, the surface disorder is more severe for an *un*passivated surface, broadening (or blurring out) the sub-band profile.[15] Passivation reduces the surface disorder, enabling the observance of sub-bands.

The length parameters *a* and *b* (Fig. 1a, where *a*=cross-sectional distance from the InAs NW core to the Si substrate and *b*=NW radius) were used to obtain the insulator (gate) capacitance per length $C_{ins}/L$ by numerical simulation. The simulation domain includes the cylindrical InAs core, 2.5nm $InO_x$ shell, and 7nm $ZrO_2$ insulator. A finite element solution is used. Photoresist is not included in the simulation domain. The Dirichlet boundary condition is used. $C_{ins}$ is obtained by fixing the electric potential at the gate-insulator interface. The dielectric constant used for InAs=15, $InO_x$=2.2, and $ZrO_2$=25. Values obtained for $C_{ins}/L$=4.8x10$^{-11}$, 5.9x10$^{-11}$, and 6.3x10$^{-11}$ F/m correspond to InAs NWs of diameter *d*=15, 27, and 32nm (post-native oxide subtraction). Figure 1b shows a scanning electron microscope (SEM) image of a representative device prior to surface passivation.

The temperature-dependent electrical behavior of back-gated InAs NW FETs with different diameters was characterized. Figure 2a demonstrates the experimental $I_{DS}$-$V_{GS}$ behavior at $V_{DS}$=10mV from 277 down to 77K for a $d$=15nm NW (channel length $L$~8 μm). Population of individual 1D sub-bands are evident as the temperature is lowered below 150K. The steps that appear in the $I_{DS}$-$V_{GS}$ curve arise from the population of 1D sub-bands as is confirmed by simulation and discussed later in the text. Figure 2b shows the experimental length normalized $I_{DS}$-$V_{GS}$ behavior at $V_{DS}$=10mV for three NWs with different diameters $d$=15, 27, and 32nm at 100K. The greater effect of quantum confinement is apparent for the smaller diameter $d$=15nm NW, with the sub-bands more pronounced and the sub-band energy spacing greater compared to the larger diameter NWs. More sub-bands contribute to the current flow for the larger diameter $d$=27 and 32nm NWs over the same $V_{GS}$ range. This is better highlighted in Supplementary Section Figure 1 where the transconductance $G_m = \frac{dI_{DS}}{dV_{GS}}\big|_{V_{DS}}$ is plotted as a function of $V_{GS}$. The $G_m$ maxima correspond to the sub-band edges. As expected, the number of $G_m$ maxima over the same $V_{GS}$ range for all three NWs increases with increasing NW diameter due to more sub-bands contributing to the current, with 3 and 4 peaks present for the 15 and 32 nm NWs, respectively, for the explored gate voltage range. The gate voltage spacing $\Delta V_{GS}$ between $G_m$ peaks decreases with increasing NW diameter. The reduction in $\Delta V_{GS}$ is most apparent between the 15nm NW and two larger diameter NWs. A noteworthy point is that the height of the first step is half of that of the second and third steps (Fig. 2). As explained in detail later, this is attributed to the two-fold degeneracy of certain sub-bands arising from the structural symmetry of cylindrical NWs.

To gain better insight into the experimental results, simulations were performed by self-consistently solving the Poisson and Schrodinger equations.[25] The single band effective mass model is used and solved in the cross-sectional plane of the NW. Spin degeneracy (2) is considered in the simulation. For the simulation, the cylindrical InAs core, 2.5nm $InO_x$ shell, and 7nm $ZrO_2$ insulator is considered. There is a finite potential barrier in between the different layers, and an infinite wall boundary condition is applied to the Schrodinger equation at the interface between $ZrO_2$ and vacuum. Figure 3a shows the simulated number of populated 1D sub-bands $N_{sub-band}$ as a function of $V_{GS}$ at $T=0K$ for three NWs with diameters $d=15$, 27, and 32nm. The diameters chosen for simulation match the experimental diameters in Figure 2b. The relationship $V_{GS} = \frac{E}{q} - \frac{Q}{C_{ins}} + V_{fb}$ was used to convert $V_{GS}$ to electron energy $E$ where $Q$ is the total charge in the NW, $q$ is the elemental charge, and $V_{fb}$ is the flatband voltage and was used as the fitting parameter. Values for $V_{fb}=2.7$, 1.2, and -0.2V for the $d=15$, 27, and 32nm NWs, respectively. The charge density at $T=0K$ is $\frac{Q}{L} = e\sum_n \frac{2\sqrt{2m(E-E_n)}}{\pi\hbar}$, for the energy $E > E_n$. Here, $E_n$ is the $n^{th}$ sub-band edge energy, and $m=0.023m_o$ is the InAs electron effective mass used in the simulations. The number of sub-bands $N_{sub-band}(E_F)$ represents the populated sub-bands with energies less than the Fermi level $E_F$. In Figure 3a, the "jump" from the first to third sub-band (and third to fifth) is a result of the two-fold degeneracy in eigenenergies where $E_2=E_3$ and $E_4=E_5$. This two-fold degeneracy explains why the height of the $2^{nd}$ and $3^{rd}$ steps in the experimental transfer characteristics is twice that of the $1^{st}$ step (Fig. 2). Figure 3b shows the temperature broadening ($T=100K$) of the calculated profile of the sub-bands population as a function of $V_{GS}$ to better match the experimental conditions. Here, the charge density for $T>0K$ is $\frac{Q_T}{L} = \frac{2e}{\pi} \int_0^\infty \sum_n \frac{1}{1+\exp\left[\left(E_n+\frac{\hbar^2 k^2}{2m}-E_F\right)/kT\right]} dk$, for $E > E_n$, where $\frac{\hbar^2 k^2}{2m}$ is the kinetic energy of an electron traveling in the longitudinal direction of the NW. The total number of populated sub-bands, when the effect of temperature (thermal broadening) has been taken into account, is $N_{sub-band,T}(E_F) = \int N_{sub-band}(E) F(E - E_F) dE$. Here, $F(E) = -\frac{\partial f(E)}{\partial E} = \frac{1}{4k_B T} sech^2\left(\frac{E}{2k_B T}\right)$ is the thermal broadening function, which takes into account the temperature broadening of the electron distribution in the sub-bands and $f(E)$ is the Fermi function. Figure 3c shows further broadening of the sub-bands population profile due to the effects of disorder at the sub-band edges (also at $T$=100K). The total number of populated sub-bands, when the combined effects of temperature (thermal broadening) and disorder (disorder broadening) have been taken into account, is $N_{sub-band,G}(E_F) = \int N_{sub-band}(E) F(E - E_F) G(E - E_F) dE$. Here, $G(E) = \frac{1}{c\sqrt{2\pi}} exp\left(\frac{-E^2}{2c^2}\right)$, the Gaussian distribution function with variance $c^2$=(0.025eV)$^2$. The combined effects of temperature and disorder broadening further clarify why the sub-bands do not appear as abrupt conductance steps in the experimental measurements.

Calculated probability density profiles for a NW with diameter $d$=15nm are shown in Figure 4 for the five lowest lateral modes (sub-bands). The eigenenergies corresponding to these modes are $E_1$=0.11 eV, $E_2$=$E_3$=0.28 eV, and $E_4$=$E_5$=0.52 eV. Using cylindrical coordinates, the Schrodinger equation can be separated into longitudinal, angular, and radial components. As a result, the eigenfunctions are separable into longitudinal, angular, and radial parts, with the angular eigenfunctions and eigenvalues containing the angular quantum number $n$, and the radial part having $n_p$ nodes along the radial direction. Here, $n$ and $n_p$ are analogous to the quantum numbers $n_x$ and $n_y$ in a square wire. The lateral modes $n_p$,+$n$ and $n_p$,-$n$ are degenerate; the linear combination of the ±$n$ eigenfunctions can be symmetric or antisymmetric with respect to reflection about the radial axis of the NW.[23,26,27] This is the reason for the two-fold degeneracy evident in lateral modes two and three ($E_2$=$E_3$) and four and five ($E_4$=$E_5$). Lateral modes one and six ($E_6$, not shown) have rotational symmetry where $n$=0, and therefore no degeneracy. These wavefunction profiles correspond to the number of populated sub-bands $N_{sub\text{-}band}$ as a function of $V_{GS}$ plots shown in Figure 3.

The simulated and experimental sub-band energies (lateral mode eigenenergies) $E_1$ to $E_5$ as a function of NW diameter for ten different NWs are shown in Figure 5, with the experimental data closely matching the simulation. The experimental energies were obtained from the peak maxima in the $G_m$ as a function of $V_{GS}$ plots (see Supplementary Section Figure 1), and relationship between electron energy $E$ and $V_{GS}$ for each NW diameter as previously discussed. The first experimentally obtained sub-band energy $E_1$ corresponds to the first $G_m$ peak, while the higher experimental sub-band energies correspond to the $G_m$ maxima. Because the threshold voltage $V_t$ can be shifted as a result of a number of factors including oxide fixed charges and electrostatics that are independent of NW diameter, the experimentally obtained $E_1$ was set to equal the simulated $E_1$ and all higher experimentally obtained sub-band energies were shifted up or down accordingly. The $G_m$ maxima over the same $V_{GS}$ range correspond to higher energy sub-bands for larger diameter nanowires as more sub-bands contribute to the current. As expected, the sub-band energy spacing (degree of quantum confinement) decreases with increasing NW diameter. The quantization for sub-30nm NWs causes the 1D sub-band spacing to be on the order of 100's of meV. The results suggest that InAs NWs with <30nm diameter can be treated as a 1D system, even at room temperature since the sub-band spacing is >>kT.

In summary, we have electrically mapped the 1D sub-bands of surface passivated InAs NWs as a function of diameter through detailed simulation and experiments. The two-fold degeneracy in certain sub-band energies arising from structural symmetry is experimentally observed for the first time. The results provide new insight into the band-structure and density of states of III-V NWs with important practical implications towards exploring the performance limits of their associated devices.

**Acknowledgement**

This work was funded by Intel, FCRP/MSD Focus Center and NSF E3S Center. The materials synthesis and characterization part of this work was partially supported by the Director, Office of Science, Office of Basic Energy Sciences, Materials Sciences and Engineering Division, of the U.S. Department of Energy under Contract No. DE-AC02-05CH11231. A.J. acknowledges a Sloan Research Fellowship, NSF CAREER Award, and support from the World Class University program at Sunchon National University. A. C. F. acknowledges an Intel Graduate Fellowship. R.K. acknowledges an NSF Graduate Fellowship. J.G. acknowledges support form NSF and SRC.

**Supporting information**

Transconductance versus gate voltage plots for different diameter NWs. This material is available free of charge via the Internet at http://pubs.acs.org.

**Figure Captions**

**Figure 1**. (a) Schematic of the cross-sectional structure of a representative back-gated cylindrical InAs NW device used in this study. (b) SEM image of a representative device prior to $ZrO_2$ and photoresist surface passivation.

**Figure 2**. (a) Experimental temperature dependent $I_{DS}$-$V_{GS}$ behavior at $V_{DS}$=10mV for a diameter $d$=15nm NW ($L$=8.07µm). Clear steps arising from the electron population of individual 1D sub-bands as a function of $V_{GS}$ are observed at <150 K. (b) Experimental length normalized $I_{DS}$-$V_{GS}$ behavior at $V_{DS}$=10mV for three NWs with $d$=15, 27, and 32nm at 100K. The $I_{DS}$-$V_{GS}$ curves for the $d$=27 and 15nm NWs are intentionally offset by 1 and 2V, respectively, for improved clarity.

**Figure 3**. (a) Simulated number of sub-bands $N_{sub\text{-}band}$ as a function of $V_{GS}$ at $T$=0K for three NWs with diameters $d$=15, 27, and 32nm. (b) Simulated temperature broadening at $T$=100K of the calculated sub-bands population profile from (a). (c) Simulated broadening due to disorder *and* temperature (at 100K) of the sub-bands population profile.

**Figure 4**. Calculated probability density profiles for a $d$=15nm NW for the five lowest lateral modes (sub-bands). A two-fold degeneracy is observed for certain sub-bands (e.g., $E_2$=$E_3$, $E_4$=$E_5$) arising from the structural symmetry of cylindrical NWs.

**Figure 5**. Simulated and experimental sub-band energies as a function of NW diameter.

# Figure 1

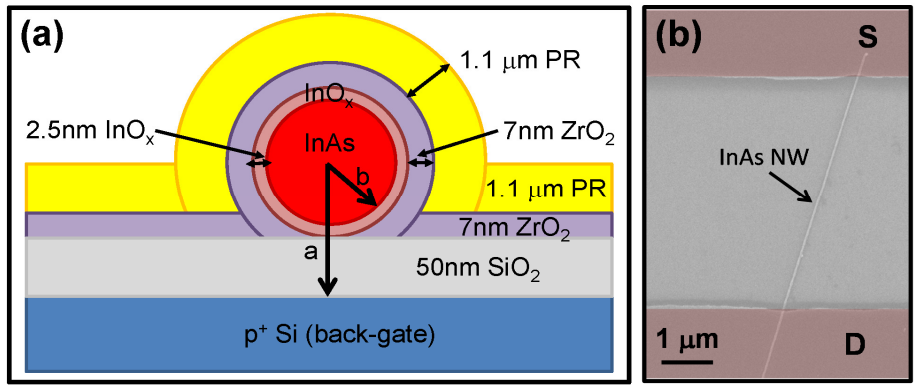

**Figure 2**

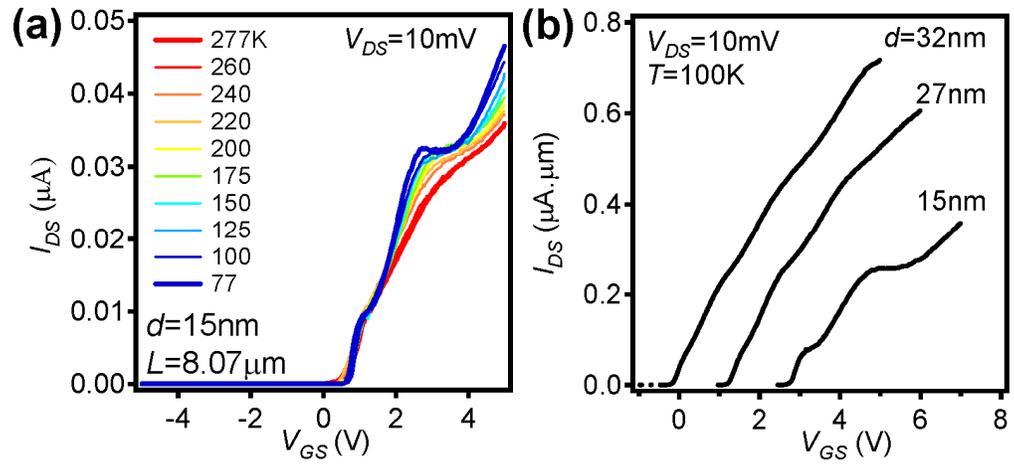

**Figure 3**

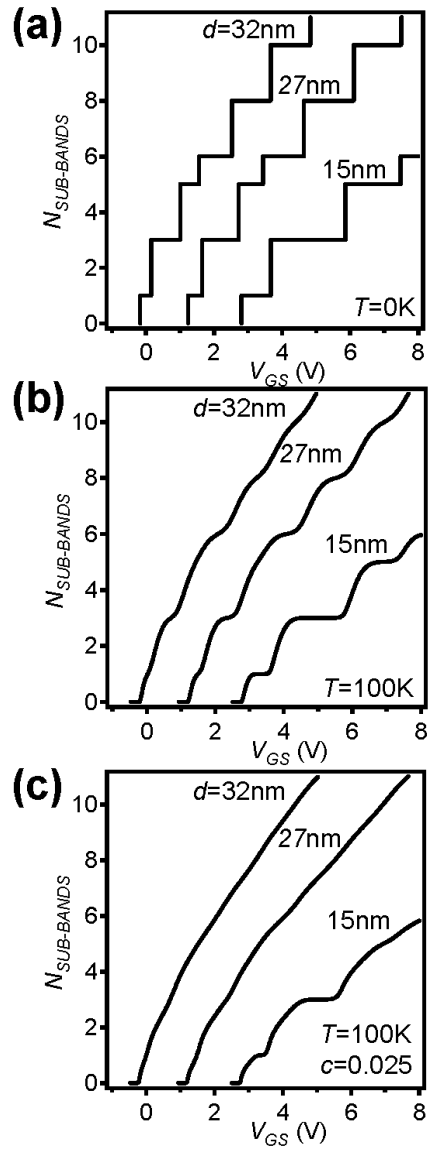

# Figure 4

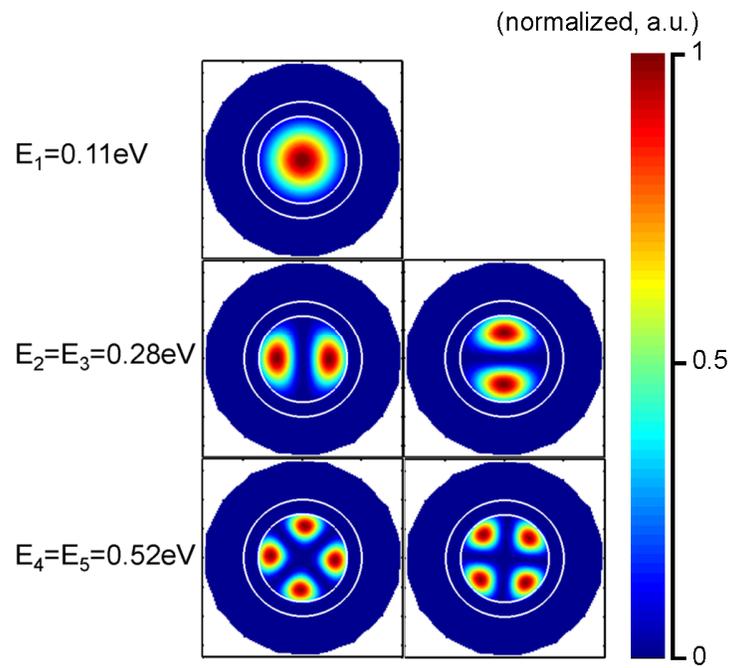

**Figure 5**

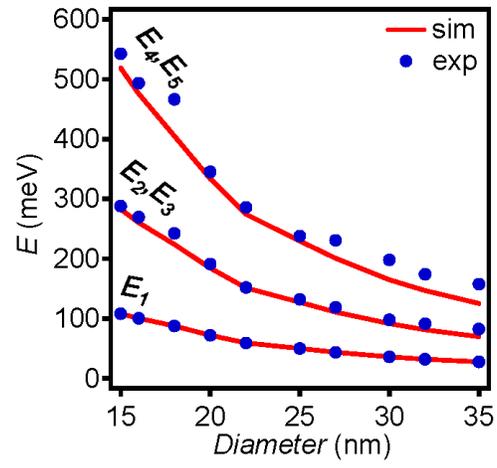

# TOC Figure

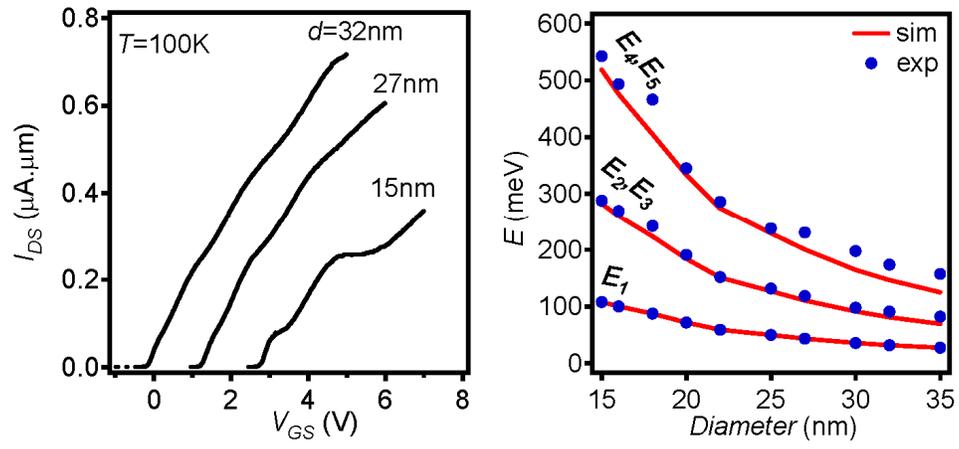

# Observation of degenerate one-dimensional sub-bands in cylindrical InAs nanowires

Alexandra C. Ford[1,2,3,†], S. Bala Kumar[4,†], Rehan Kapadia[1,2,3], Jing Guo[4], Ali Javey[1,2,3,*]

[1] Electrical Engineering and Computer Sciences, University of California, Berkeley, CA, 94720, USA.

[2] Materials Sciences Division, Lawrence Berkeley National Laboratory, Berkeley, CA 94720, USA.

[3] Berkeley Sensor and Actuator Center, University of California, Berkeley, CA, 94720, USA.

[4] Electrical and Computer Engineering, University of Florida, Gainesville, Florida 32611

[†]These authors contributed equally.

* Corresponding author: ajavey@eecs.berkeley.edu

**SUPPORTING INFORMATION**

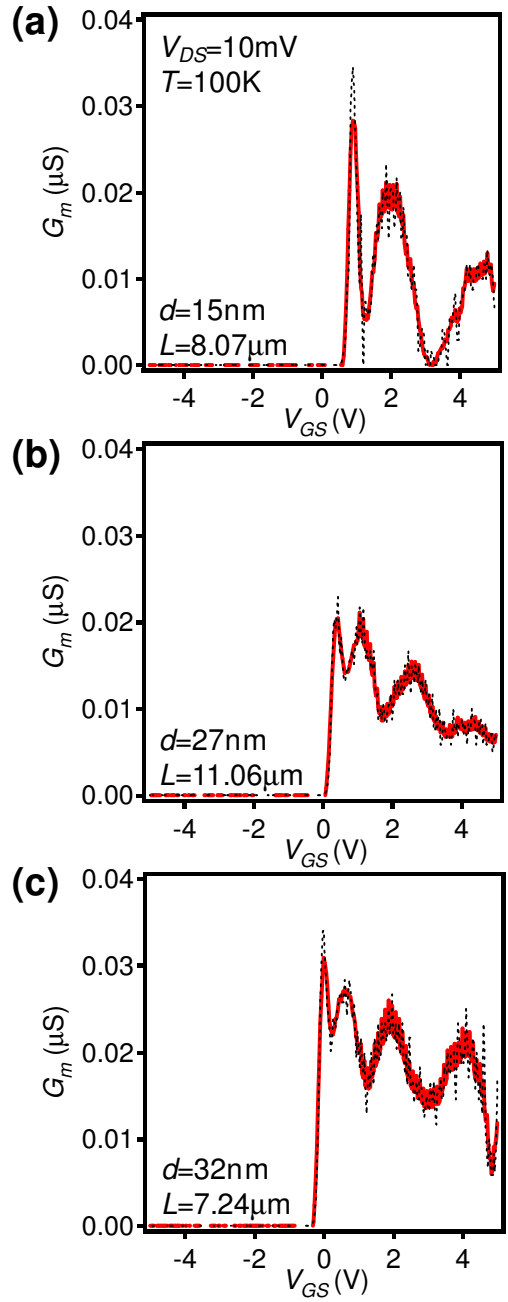

**Figure S1.** Transconductance, $G_m$, vs. gate voltage, $V_{GS}$, for three InAs NW FETs of diameter (a) 15nm, (b) 27nm, and (c) 32nm. The plots correspond to the same NWs shown in Figure 2b. The experimental energies plotted in Figure 5 were found from the maxima of $G_m$-$V_{GS}$ plots.